\documentclass[aps,preprint,nofootinbib]{revtex4-2}

\usepackage{graphicx}
\usepackage{amsmath}
\usepackage{amsfonts}
\usepackage{amssymb}
\usepackage{bm}
\usepackage{xcolor}
\usepackage{ulem}

\begin{document}

\title{Charged particle energization by low-amplitude electrostatic waves at cyclotron harmonics}

\author{Fabio Sattin}
\email{fabio.sattin@igi.cnr.it}
\affiliation{Consorzio RFX (CNR, ENEA, INFN, Università di Padova, 
Acciaierie Venete SpA), Corso Stati Uniti 4, 35127 Padova, Italy}

\author{Lorenzo Martinelli}
\affiliation{Department of Energy, Politecnico di Milano, Via Lambruschini, 20156 Milano, Italy}

\begin{abstract}
The system made by a charged particle interacting with a single electrostatic wave which propagates perpendicularly to the magnetic field, at a frequency larger than the cyclotron one, has been extensively studied in literature due to its implications with ion heating in magnetized plasmas. It is known that a threshold in the electrostatic potential must be exceeded in order for stochastic particle motion and heating to occur. Regardless its amplitude, however, the electrostatic wave induces a periodic oscillation in the particle motion. We show, by analytical and  numerical arguments, that this dynamics is non-adiabatic, meaning that the particle does not land back to its initial state when the wave is slowly turned off. This way, particle energization (although, not rigorously heating) occurs even under sub-threshold conditions.

\hfill


\end{abstract}

\maketitle

\section{Introduction}

The interaction between charged particles and electrostatic waves is an old topic in plasma physics-- both astrophysical and laboratory ones--due to its great importance for ion heating. \\
In this paper, we will address the specific case of a particle interacting with a wave train propagating transversely to the magnetic field, with frequency larger than the particle Larmor one: it is the case, for example, of a lower-hybrid wave interacting with ions. 
This problem was first studied by Smith and Kaufman \cite{smith75} (for wave propagation oblique to the magnetic field) and Fukuyama {\it et al} \cite{fukuyama77} (for perpendicular propagation) and, a few years later, revisited by Chia {\it et al} \cite{chia96}. The conclusion is that particle heating is due to nonlinear mechanisms: there is an irreversible flow of energy to the particle, with stochastization of orbits and diffusion in energy, provided that the wave amplitude exceeds a minimum value, beyond which overlap of different resonant islands in phase space occurs . The half-width of the $l$-th island in the particle velocity space is given by \cite{smith75,benisti98} $w = \sqrt{4 \, e/m \, \phi_0 \, J_l(k_\perp \rho)}$, where $e,m$ are the particle's charge and mass, $\phi_0$ the amplitude of the electrostatic fluctuation, $\rho$ the thermal Larmor radius, $k_\perp$ the wavenumber perpendicular to the field, $J$ the Bessel function of the first kind, and its order $l$ is an integer number $l = 0, \pm 1, \pm 2, ...$. According to the Chirikov criterion, overlap between neighbouring islands occurs when the sum of their half-widths is of order or greater than the distance between their centers, which are defined in terms of the parallel velocity by the relation $\omega - k_z \, v_l = l \, \Omega$, where $\omega, \Omega$ are respectively the wave angular frequency and the cyclotron frequency, $k_z$ is the wavenumber parallel to the field, and $v_l$ is the particle's parallel velocity. At small amplitude $\phi_0$, conversely, one may demonstrate that approximate constants of the motion do exist, ensuring regular and smooth motion. Accordingly, no irreversible transfer of energy and heating is expected in this case. The necessity of a minimal wave amplitude does severely constrain the possibility of particle heating by monochromatic spectra; thus scholars have often attempted to workaround the problem by invoking the interaction not with single waves but with a broadband spectrum \cite{benisti98, jorns13}.  \\
The purpose of this  short paper is to revisit this problem. We will point out that an irreversible flow of energy from the wave to the particle does actually take place regardless the amplitude of the wave. However, it is different from a continuous heating, though, in the sense that the particle energy does not grow indefinitely.  
We will show, using the very same analysis employed in the papers cited above, that the energy of a particle interacting with a wave train, eventually does not land to its pristine value, but rather may range between fixed lower and upper bounds that do not depend upon the wave amplitude. 
If the particle is chosen with an initial small energy, the consequence is its effective energization.      

\section{Hamiltonian analysis of the wave-particle system}

We will consider the case of a charged particle in a uniform magnetic field aligned along the z axis, and a plane electrostatic wave propagating along the x axis. No feedback of the particle upon the wave is considered.
In dimensionless form, where (particle mass) = (particle charge) = (wavenumber) = (Larmor frequency) = 1,  the hamiltonian is written
\begin{equation}
H = \frac{p^2}{2} +\frac{x^2}{2} + A \cos \left( x - \omega \, t \right)
\label{eq:hpq}
\end{equation}
The wave frequency is close to an harmonic of the Larmor frequency, i.e. $\omega \approx M$, with $M$ integer $>1$. \\
We make a first transformation of (\ref{eq:hpq}) to action-angle variables $(I, \theta)$: $x = \sqrt{2 I} \sin (\theta), p = \sqrt{2 I} \cos (\theta)$. In order to simplify the notation, the auxiliary variable $r = \sqrt{2 I}$ will also be employed. We arrive thus to 
\begin{equation}
H = I + A \cos \left( r \sin(\theta) - \omega \, t \right)
\label{eq:hit}
\end{equation}
We apply now the Jacobi--Anger expansion: 
\begin{equation}
\cos \left( r \sin (\theta) - \omega \, t \right) = \sum_{m = -\infty}^{\infty} J_m (r) \cos (m \, \theta - \omega \,t) , 
\end{equation}
where $J_m$ is the Bessel function of the first kind of order $m$,  and arrive to
\begin{equation}
H = I + A \sum_{m = -\infty}^{\infty} J_m (r) \cos (m \, \theta - \omega \,t)
\label{eq:hit2}
\end{equation} 

\subsection{Exact resonance case}
\begin{figure}
\includegraphics[width=10.5cm]{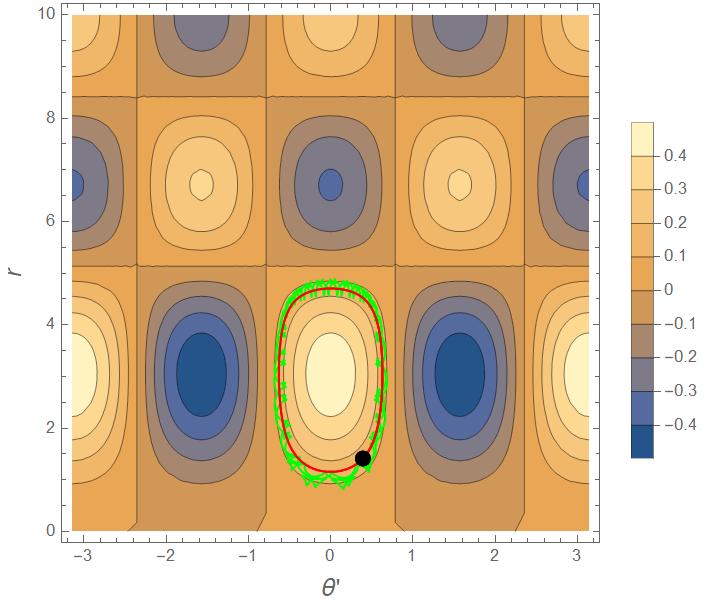}
\caption{Contour plot of $J_M (r) \cos(M \theta') $ for $M = 2$. The red and green curves are respectively the solutions of the truncated and full Hamiltonian equations. Throughout this paper, by full Hamiltonian we will mean that retain all terms within the interval $(M-7, M+7)$. The starting position is marked by the black dot. The wave amplitude is $A = 1/3$. }
\label{fig:uno}
\end{figure}

At this point, we make the hypothesis that the wave frequency is an exact multiple of the Larmor frequency $\omega = M$  (later, we will relax this assumption).  
We move to the frame rotating with the Larmor frequency, employing the generating function $F = I (\theta - t)$. The new Hamiltonian writes
\begin{equation}
H' = A \sum_{m = -\infty}^{\infty} J_m (r) \cos (m \, \theta' - (\omega -m) \,t)
\label{eq:h1}
\end{equation}
We split $H'$ into two contributions:
\begin{equation}
H' = A J_M (r) \cos(M \theta') + A \sum_{m\neq M} J_m (r) \cos (m \,\theta' - (M-m) \, t )
\label{eq:hf}
\end{equation}
The second, time-dependent term, provides only a perturbation to the first one, and may be removed by making a temporal average over one Larmor period (Krylov--Bogoliubov averaging). We will validate later, numerically, the goodness of this simplification. We are thus left with the first term alone, and with the corresponding Hamilton equations:
\begin{eqnarray}
\frac{dI}{dt} &=& A\,M \,J_M (r) \sin(M \theta')  \label{eq:didt} \\
\frac{d\theta}{dt} &=& A \frac{J_{M-1}(r) - J_{M+1} (r)}{2 \,r} \cos (M \,  \theta')
\label{eq:hameq}
\end{eqnarray}
 From the inspection of these equations, we establish that: (i) When $A$ is strictly constant, we may rescale time: $t' = A \, t$, so that $A$ disappears from the equations of motion. The solutions depend only trivially from the wave amplitude (provided that it is non-zero) which plays just the role of a time scaling factor. (ii) Under the same assumption of constant $A$, the system is autonomous and one-dimensional, thus its orbits are automatically regular. (iii) Along the $I$ direction, the motion is bounded between two successive zeros of the Bessel function: $J_M (r) = 0$. This, too, is true irrespective of the value of $A$. The solutions $(I, \theta')$ of the above system are functions rotating periodically along contours of constant $H'$. In Fig. (\ref{fig:uno}) we provide an example of such motion, with superimposed the full solution: i.e., retaining even the non-resonant terms in the hamiltonian $H'$. 
 
\begin{figure}
\includegraphics[width=10.5cm]{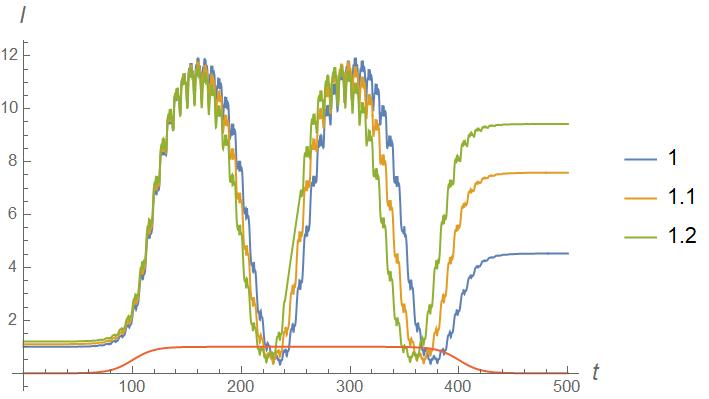}
\caption{Time traces of $I(t)$ computed using the full Hamiltonian (\ref{eq:hf}) with $M = 2$. The blue, orange and green curves correspond to $I(0)$ as given in the legend, while the initial phase is $\theta'(0) = \pi/8$ for all. The initial conditions, thus, have been chosen close to the bottom of the ellipsoidal curve tracked by the green trajectory in Figure (1). The wave amplitude is $A = 1/3 f(t)$, with $f(t)$ defined in Eq. (\ref{eq:shape}), and plotted in the figure as a red curve. Its parameters are $dt = 20, t_s = 100, t_f = 400$. Notice that, with this choice of parameters, the time needed for $A$ to grow from zero to its flattop value is $\approx 130$ time units, equivalent to about 20 Larmor periods.}
\label{fig:due}
\end{figure}

\begin{figure}
\includegraphics[width=10.5cm]{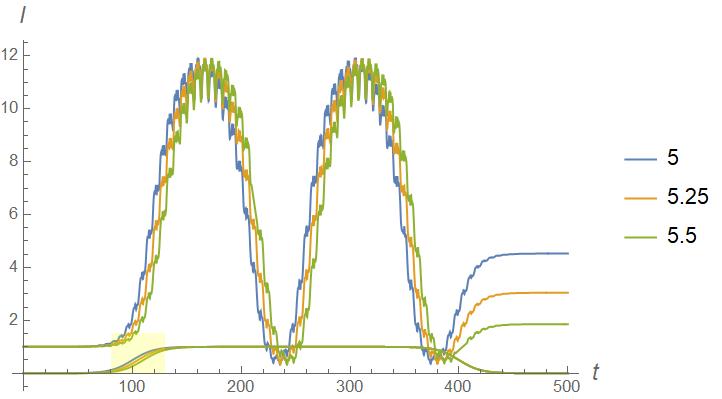}
\caption{This figure has the same content of Figure (\ref{fig:due}). In this case, the initial conditions are the same for all trajectories $( I(0) = 1, \theta'(0) = \pi/8)$, but the shape function grows over very slightly different time scales between runs: $t_s/dt = 5,5.25, 5.5$ for the three cases. The yellow rectangle centered at $t \approx 100, I \approx 1)$ marks the region where $f$ differs between runs. All other parameters are the same from the previous figure. }
\label{fig:tre}
\end{figure}

Let us now consider the physically realistic situation of a wave train of finite duration. That is, the wave amplitude is first very slowly switched on and then, after some time, turned off in an equally slow manner. Throughout this paper, $A$ will be  modulated by the following shape  function:
\begin{equation}
f(t) = \frac{1}{4} \,\left(1+\tanh\left[\frac{(t - t_s)}{dt}\right]\right) \left(1+ \tanh\left[\frac{(t_f - t)}{dt}\right]\right)
\label{eq:shape}
\end{equation}
That is, $A \to A f(t)$. The shape function $f$ is designed so as to be approximately equal to unity inside the interval $(t_s,t_f)$, and zero outside of it, with a transition width of order $dt$. \\
For a regular system one expects a continuous adjustment to these adiabatically varying conditions. In particular, the system is expected to return to its initial energy once the perturbation is switched off. Actually, this is certainly true for $H'$, which is proportional to $A$. However, $H'$ does not stand for the particle energy. In the laboratory frame, the latter is instead given by $H$ of Eq. (\ref{eq:hit}) or (\ref{eq:hit2}): in order to check the (non)adiabaticity of the motion, it is therefore $I$ the dynamical variable to be tracked. We have seen that $I$ evolves periodically, with a period that is proportional to $1/A$. When the wave is switched off, the action remains frozen amid the two successive zeros of $dI/dt$ in Eq. (\ref{eq:didt}), i.e. when $J_M(r) = 0$, at a value which is exactly determined, once its initial conditions and the parameters of the wave train are given. Any ignorance about their precise values results in a sensitive (although not exponential) indeterminacy about the final value of $I$, $I_{fin}$. In actual situations, initial conditions are not known, but picked up from some statistical distribution, and the wave train parameters are not always finely tunable by the experimentalist, thus $I_{fin}$ turns out to be effectively a random quantity.  \\
In Figs. (\ref{fig:due},\ref{fig:tre}) we provide a numerical confirmation of this claim. In the first plot, we display the time traces of $I$ for the very same wave parameters, but starting from close conditions. The three trajectories perform two complete rotations in the action-angle phase space then, while they are one their way along the third rotation, the wave is switched off at $t \approx 400$, freezing the action at the corresponding value. It is appreciable how the three trajectories have widely different conclusions even though exponential separation of trajectories appropriate to chaotic systems is not at work here.

 Figure (\ref{fig:tre}) shows the complementary case: there, we start the particle trajectory from one and the same initial condition, but allow for very tiny changes to the duration of the wave train: the starting time is very slightly delayed between runs, as evidenced by the pale yellow rectangle in Figure (3). Here, too, we observe substantially different values of the action once it is frozen.    

\subsection{Off-resonance case}
\begin{figure}
\includegraphics[width=10.5cm]{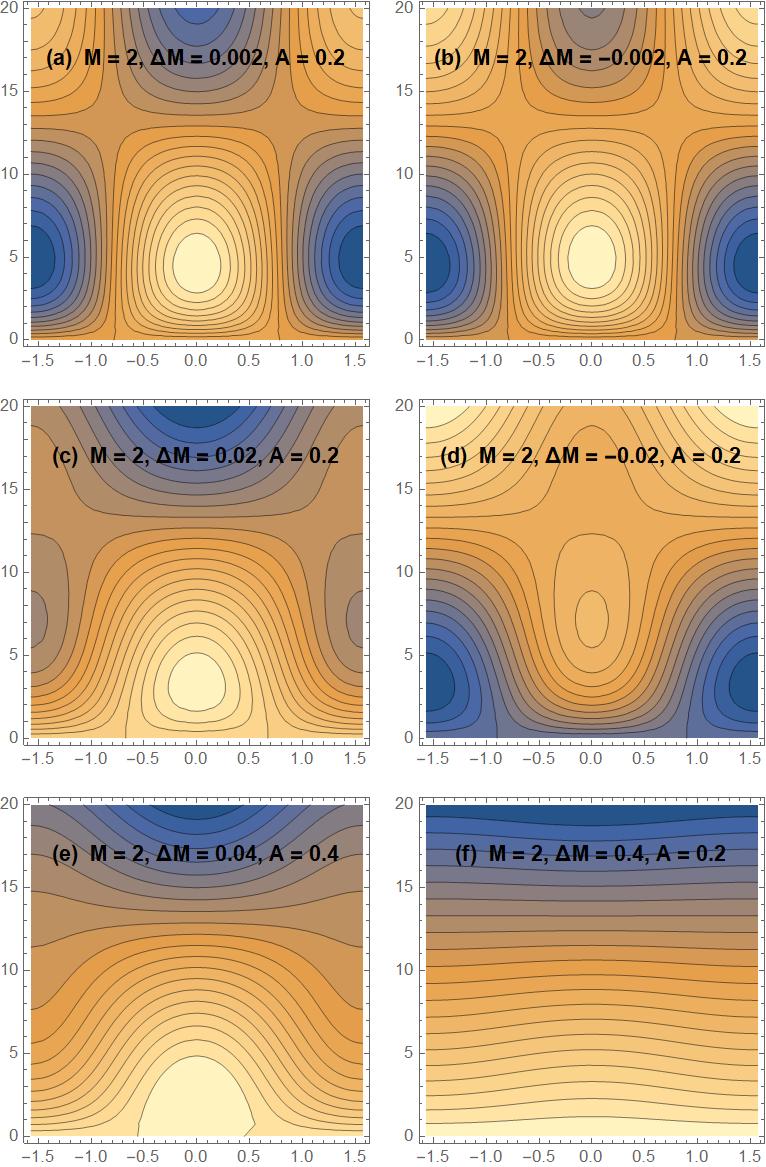}
\caption{Contour plots of $H'$ (Eq. \ref{eq:ptr}) for different combinations of the parameters $\Delta M, M, A$.  }
\label{fig:quattro}
\end{figure}

\begin{figure}
\includegraphics[width=10.5cm]{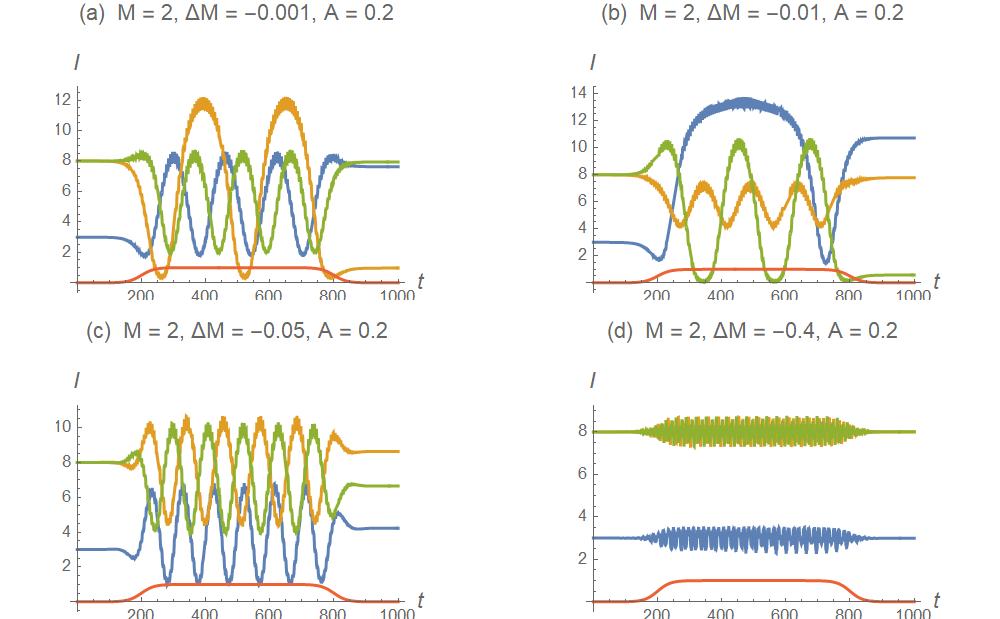}
\caption{Examples of trajectories for different sets of parameters and different initial conditions. In all subplots, the initial conditions are the same: $I(0) = 3, \theta(0) = -1.3$ (blue curve); $I(0) = 8, \theta(0) = -1.0$ (orange curve); $I(0) = 8, \theta(0) = 0.1$ (green curve). The red curve is the shape function $f$ modulating the amplitude $A$, like in Fig. (\ref{fig:due}). The parameters $M, \Delta M, A$ employed for each subplot are reported as titles of the subplot itself.  All numerical trajectories are computed using the full hamiltonian (\ref{eq:hf}), not their truncated versions. }
\label{fig:cinque}
\end{figure}

What we may expect in the more general and realistic case, when the wave frequency is not an exact multiple of Larmor frequency? Let us write $\omega = M + \delta M$ with $0 < \delta M < 1$,  and apply to Eq. (\ref{eq:hit2}) the canonical transformation generated by the function $F' = I \, \left( \theta - (1+ \delta M/M) t \right)$. \\
The Hamiltonian becomes 
\begin{equation}
H = P + T + R \, ,  \,
P =  - \frac{\delta M}{M} I  \; , \quad  T = A \, J_M(r) \cos (M \, \theta')
\label{eq:ptr}
\end{equation}
In $R$ we have included all the terms with $m \neq M$, which still contain a time dependence, and cancel out after time averaging. Hence, we will set $R = 0$ from now on. \\
The system behaves in opposite ways when either $P$ or $T$ dominate within the Hamiltonian. When $P$ dominates, the system reduces to a harmonic oscillator, in action-angle coordinates. All orbits are passing, with $I$ = constant and $\dot{\theta} = -\delta M/M$. The addition of the small perturbation $T$ does not impact upon the qualitative behavior of the system, that remains adiabatic.  Conversely, when $T$ dominates, we are back to a situation analogous to the exact-resonance one, all orbits are closed, and the regime is fully non-adiabatic. In the intermediate situation, we have a mixture of open and closed orbits. If we restrict ourselves to moderate values of $I$: say, $I$ between zero and the first root of the Bessel function, the balance between the terms $P$ and $T$ is roughly ruled by the ratio $(|\Delta M|/M) \, A^{-1}$.  In Figure (\ref{fig:quattro}) we show the contour plots of the Hamiltonian (\ref{eq:ptr}) for different values of this ratio, highlighting the change in its topology. \\
In Figure (\ref{fig:cinque}), finally, we plot a few instances of trajectories pertaining to the different regimes, highlighting the transition from the adiabatic to the non-adiabatic behavior. \footnote{It is worth mentioning that the case with $M = 2, \Delta M \approx 1/2$ is non-generic since, with this specific choice of parameters, {\it two} terms stand out in Eq. (\ref{eq:hit2}), having approximately the same weight: those with $m = M$, and $m = M+1$. The interference between the two terms produce open trajectories--and therefore nonadiabatic behaviour--in the landscape of $H$. By increasing $M$ or slightly altering $\Delta M$ the effect is immediately lost. } 

\section{Conclusions}
It is commonly acknowledged that regular integrable dynamics in most situations entails adiabatic behaviour.
Exceptions to this rule are well  known, usually involve the existence of one or more separatrices that separate the phase-space into disjoint basins, which may be connected to each other by small perturbations. Some examples are, in the classical domain, a magnetized particle interacting with low-frequency waves \cite{escande19,sattin23} and, in the quantal domain, the Landau-Zener model: in both cases, the system is close to integrable during most of the time, only when the trajectory approaches the separatrix the irregularity of the orbit appears clearly. A generic analysis, encompassing both classical and quantum aspects, is found in \cite{zhang13}. 
The present case is completely different from those well-known scenarios, though, since it does not involve any separatrix crossing. In our case, turning the wave on induces a rotation (superposed to the cyclotron rotation) of the particle. This dynamics is not adiabatic, in the sense that turning off the wave stops the rotation, but the particle coordinates (and its energy) remain frozen at a value different from the starting one.    
Since the final energy acquired by the particle is bounded from above, there cannot be true heating but one still observes the spreading of any initially monoenergetic distribution. Furthermore, if the initial particle is very cold--that is, $I(0)$ is much smaller than the first zero of the Bessel function $J_M$, then the final energy, with large probability, will be larger than $I(0)$, effectively corresponding to an energization. In order to have a glimpse about the typical figures involved, we recall that the dimensionless action $I$ is converted to energy in electron-volts after multiplication by $E_0 = e^{-1} \, m \, \Omega^2 \, k^{-2}$, where $e$ is the elementary charge, $m$ the particle mass, $\Omega$ its Larmor frequency, and $k$ the wavenumber (All quantities are in MKS units). If we express $k$ in terms of the the inverse of the particle thermal Larmor radius:  $k  = q \, \rho^{-1}$, then $E_0$ turns out to be proportional to the ion temperature $T: E_0 = T \, q^{-2} $, and the energy variation is $\Delta E = E_0 (I_{fin}- I_{ini})$.
Since $I_{fin}-I_{ini}$ may range in the order of tens  (in our normalized units, the first root of the Bessel function varies between $\approx 13.2$ and $\approx 38$ when $M$ ranges between 2 and 5), waves of small amplitude whose wavelength is close to the Larmor radius (i.e., $ q \approx O(1)$) may deliver a significant contribution to the ion thermal content.

\end{document}